# Cavity-enhanced optical readout of a single solid-state spin

Shuo Sun,[1] Hyochul Kim,[1] Glenn S. Solomon,[2] and Edo Waks[1,a]

[1]Department of Electrical and Computer Engineering, Institute for Research in Electronics and Applied Physics, and Joint Quantum Institute, University of Maryland, College Park, Maryland, 20742, USA

[2]Joint Quantum Institute, University of Maryland and National Institute of Standards and Technology, College Park, Maryland, 20742, USA

**Abstract**

We demonstrate optical readout of a single spin using cavity quantum electrodynamics. The spin is based on a single trapped electron in a quantum dot that has a poor branching ratio of 0.43. Selectively coupling one of the optical transitions of the quantum dot to the cavity mode results in a spin-dependent cavity reflectivity that enables spin readout by monitoring the reflected intensity of an incident optical field. Using this approach, we demonstrate spin readout fidelity of 0.61. Achieving this fidelity using resonance fluorescence from a bare dot would require 43 times improvement in photon collection efficiency.

[a] Email: edowaks@umd.edu.

## I. Introduction

Spins in solids are promising qubit systems for quantum information applications due to their scalability and prospects for developing compact chip-integrated devices[1,2]. Scalable quantum technology requires methods to measure these spins with high speed and fidelity[3]. Optical spin readout provides one of the fastest and most precise measurement methods[4,5], and is thus highly desirable for scalable quantum information processing.

Optical readout approaches typically rely on resonance fluorescence[6-10], resonance absorption[11], or optical Kerr or Faraday rotations[12-14]. The readout fidelity of these approaches is limited by the branching ratio of the spin system, defined as the probability that an optical excitation induces a spin-flip due to an undesired decay channel[15]. For example, for resonance fluorescence spectroscopy, the branching ratio determines the number of photons generated by the cycling transition before the measurement induces a spin-flip. Many confined spin systems such as quantum dot spins[16], fluorine impurities[17], and silicon-vacancy centers in diamond[18,19], do not possess a good branching ratio due to non-radiative decay mechanisms or poor selection rules. In addition, the external magnetic field direction to achieve optimal branching ratio for these confined spin systems typically conflict with the condition that allows coherent optical spin manipulations[20-23]. These qubit systems therefore require new methods for readout.

Optical cavities can significantly improve qubit readout. For example, cavities can enable quantum non-demolition measurements of the hyperfine states of single atoms by probing absorption without scattering the atom out of the trap[24-26], thereby preserving its quantum state. In solid-state systems, planar distributed Bragg reflector cavities showed impressive spin-readout fidelity at the single-shot level[10]. The cavity utilized in this work served to facilitate extraction of photons from the substrate, but did not exhibit strong light-matter coupling in the form of a Purcell effect due to a low cavity quality factor and high mode-volumes. More recent theoretical work showed that cavities operating in the high-cooperativity regime where light-matter interactions are strong can enable high-fidelity spin readout[27-29], even when the qubit has a poor branching ratio[30]. In this approach, the cavity directly modifies the radiative properties of the spin transition, while the emitter induces a spin-dependent reflectance or transmittance of a cavity that efficiently couples to an external readout laser[31-33]. This strong coupling of light to matter fundamentally improves the readout fidelity beyond the limits imposed by the branching ratios of the bare system, enabling high-fidelity spin readout in a broad range of physical systems lacking an appropriate cycling transitions. However, such cavity-enhanced spin readout remains to be experimentally demonstrated.

In this paper, we demonstrate enhanced optical readout of a single solid-state spin using cavity quantum electrodynamics. We demonstrate this spin readout approach using a spin contained in a single InAs quantum dot coupled to a photonic crystal cavity. Selectively coupling one of the optical transitions of the quantum dot to the cavity mode results in a spin-dependent cavity reflectivity that enables spin readout by monitoring the reflected intensity of an incident optical field. Using this approach, we demonstrate spin readout fidelity of 0.61. Achieving this fidelity using resonance fluorescence from a bare dot would require 43 times improvement in photon collection efficiency.

## II. Protocol for cavity-enhanced spin readout

Figure 1(a) shows the level structure of the charged quantum dot, which is composed of two ground states corresponding to spin states of the electron, denoted $|\uparrow\rangle$ and $|\downarrow\rangle$, and two excited trion states composed of two electrons and a hole, which we denote as $|\Uparrow\rangle$ and $|\Downarrow\rangle$ to highlight the spin of the hole. The spin-conserving transitions $|\Uparrow\rangle \to |\uparrow\rangle$ and $|\Downarrow\rangle \to |\downarrow\rangle$ are optically allowed, denoted as $\sigma_\uparrow$ and $\sigma_\downarrow$ in the figure. The cross-transitions $|\Uparrow\rangle \to |\downarrow\rangle$ and $|\Downarrow\rangle \to |\uparrow\rangle$ are forbidden for excitons composed purely of heavy holes, but heavy-hole light-hole mixing will render these transitions partially allowed[34]. A magnetic field applied along the growth direction (Faraday geometry) breaks the degeneracy of the optical transitions, but does not significantly alter the selection rules[16]. We define the branching ratio of a quantum dot as $R_B = \gamma/(\Gamma + \gamma)$ [15], where $\Gamma$ is the spontaneous emission rate of the optically allowed transition $|\Uparrow\rangle \to |\uparrow\rangle$ for a bare quantum dot, and $\gamma$ is the decay rate of transition $|\Uparrow\rangle \to |\downarrow\rangle$.

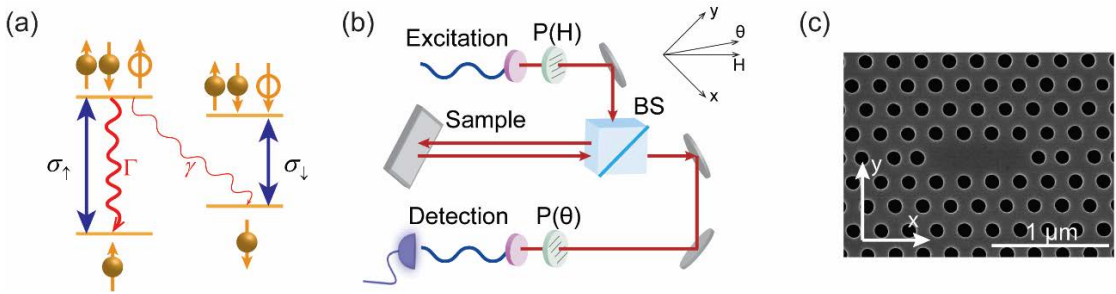

FIG. 1. (a) Energy level structure of a charged quantum dot in the presence of a magnetic field applied in the Faraday geometry. (b) Schematic setup for optical spin readout based on cavity quantum electrodynamics. BS: beam splitter; P(H): polarizer along $H$ direction; P($\theta$), polarizer along $\theta$ direction. (c) Scanning electron microscope image of a fabricated photonic crystal cavity device.

We follow the spin readout method described and analyzed in Ref. [30]. In our implementations, transition $\sigma_\uparrow$ resonantly couples with a single-sided optical cavity, whereas transition $\sigma_\downarrow$ is decoupled due to a large detuning induced by an external magnetic field. For a single incident photon that is resonant with the cavity, the cavity reflection coefficients in the cases of spin-up and spin-down states are given by $r_\uparrow = 1 - \frac{2\alpha}{C+1}$ and $r_\downarrow = 1 - 2\alpha$ respectively[31-33], where $C = 2g^2/\kappa\Gamma_d$ is the atomic cooperativity, and $\alpha = \kappa_{ex}/\kappa$ is the interference contrast. In these expressions, $g$ is the coupling strength between transition $\sigma_\uparrow$ and the cavity mode, $\kappa$ is the energy decay rate of the cavity, $\Gamma_d = \frac{\Gamma + \gamma}{2} + \gamma_d$ is the dipole decay rate of transition $\sigma_\uparrow$ where $\gamma_d$ is the dipole decoherence rate, $\kappa_{ex}$ is the cavity energy decay rate to the reflected mode. In the ideal limit of high interference contrast ($\alpha = 1$) and high cooperativity ($C \gg 1$), the reflection coefficients become $r_\uparrow = 1$ and $r_\downarrow = -1$. Thus, a photon picks up a spin-dependent $\pi$ phase shift upon reflection, which distinguishes the two spin states. Non-ideal cooperativity and interference contrast will degrade the amplitudes of the coefficients but will still lead to a change of phase shift provided $\alpha > 0.5$ and $C > 1$.

In order to convert the spin-dependent phase shift to an optical signal that performs readout, we use the polarization interferometry setup illustrated in Fig. 1(b). We inject a weak coherent field whose polarization is oriented at a 45-degree angle relative to the cavity polarization axis. The polarization component that is along the cavity is reflected with a spin-dependent phase shift, whereas the orthogonal polarization component is directly reflected from the sample surface with no phase shift. We send the reflected field to a polarizer rotated at an angle $\theta$ relative to the cavity polarization axis and focus it onto a single-mode fiber. A single-photon detector monitors the field intensity to perform spin readout. In Supplementary Material Section 1[35], we show that by properly selecting the angle $\theta$, we attain a collection probability of $P = \frac{1}{4}\beta|1+r_{\uparrow,\downarrow}|^2$ for a cavity-coupled incident photon, where $\beta$ is the coupling efficiency from the cavity spatial mode to the collection fiber. Therefore, the detector will not detect any photons when the spin is in the spin-down state ($r_\downarrow = -1$), but will detect a bright photon flux for spin-up state ($r_\uparrow = 1$). The system implements a spin-readout in an analogous way to resonance fluorescence spectroscopy.

The spin readout fidelity is limited by the number of photons reflected into the detection polarization basis before a spin-flip event occurs. If we use resonance fluorescence from a bare dot to measure the spin, the maximum number of photons we can extract is $N' = (1-R_B)/R_B$. The number of reflected photons using the cavity quantum electrodynamics approach is instead given by $N = \frac{2g^2}{\kappa\Gamma}N'$ (see Supplementary Material Section 2[35]). In photonic crystal cavities the enhancement factor $\frac{2g^2}{\kappa\Gamma}$ can be as high as 1300[36], which could correspond to three orders of magnitude improvement in the number of detected photons.

### III. Device design, fabrication, and characterization

We couple the quantum dot with a photonic crystal cavity. Figure 1(c) shows the scanning electron microscope image of the fabricated photonic crystal cavity. The initial wafer for device fabrication is composed of a 160-nm thick GaAs membrane with a single layer of InAs quantum dots at its center (density of 10-50/µm$^2$). A fraction of quantum dots in the sample are naturally charged due to residual doping background. We use a weak white light illumination to stabilize the extra electron confined in the dot. The membrane layer is grown on top of a 900-nm thick $Al_{0.78}Ga_{0.22}As$ sacrificial layer. A distributed Bragg reflector composed of 10 layers of GaAs and AlAs is grown below the sacrificial layer and acts as a high reflectivity mirror, creating a one-sided cavity. Photonic crystal structures are defined using electron-beam lithography, followed by inductively coupled plasma dry etching and selective wet etching of the sacrificial AlGaAs layer. The cavity is composed of a three-hole defect in a triangular photonic crystal with a lattice constant of 240 nm and a hole radius of 72 nm, where we shift the inner three holes adjacent to the defect to optimize the quality factor[37]. The cavity supports a small mode volume of $0.7(\lambda/n)^3$ [38], where $\lambda$ is the cavity resonant wavelength and $n$ is the refractive index of the GaAs substrate.

To optically characterize the device, we mount the sample in a closed-cycle cryostat that cools the sample to 3.6 K. An integrated superconducting magnet system applies a magnetic field of up to 9.2 T in the out-of-plane (Faraday) configuration. We excite the sample and collect the reflected signal using a confocal microscope with an objective lens that has a numerical aperture of 0.82. A single mode fiber spatially filters the collected signal to remove spurious surface reflection. We perform spectral measurements using a grating spectrometer with a spectral resolution of 7 GHz. To measure the temporal properties of the signal we perform photon counting measurements using a Single-Photon Counting Module (SPCM-NIR-14) with a time resolution of 800 ps.

We estimate an overall photon detection efficiency of our system to be 0.41%, which includes the collection efficiency of the objective lens (4.5%), transmission efficiency for a 90/10 beam splitter (90%), a fiber connector (73%), and a fiber Fabry-Perot tunable filter (40%), and the quantum efficiency of the detector (35%).

We first characterize the device by performing reflectivity measurements using a broadband LED[39]. We set the detection polarization to be orthogonal to the input field. Figure 2(a) shows the reflection spectrum as a function of magnetic field. At 0 T, the spectrum shows a bright peak due to the cavity (labeled as CM) and a second peak due to the quantum dot (labeled as QD), which is red-detuned from the cavity resonance by 0.27 nm (94 GHz). At higher magnetic field the quantum dot splits into two peaks, corresponding to the $\sigma_\uparrow$ and $\sigma_\downarrow$ transitions shown in Fig. 1(a). Measurements with a magnetic field applied in the Voigt configuration verifies that the quantum dot is charged (see Supplementary Material Section 3.1[35]).

To set the polarization analyzer to the optimal orientation for spin readout, we set the magnetic field to 0 T so that the dot is highly detuned from the cavity. We then orient the polarization analyzer to minimize the measured field intensity at the cavity resonance, which results in the cavity spectrum shown as red diamonds in Fig. 2(b). We obtain this spectrum by scanning the frequency of a tunable narrow bandwidth (< 300 kHz) laser and monitor its reflected intensity at each frequency. We then increase the magnetic field to 3.7 T where transition $\sigma_\uparrow$ is resonant with the cavity mode. We also introduce a second narrow linewidth laser resonant with the $\sigma_\downarrow$ transition to optically pump the spin to the spin-up state[40]. The blue circles in Fig. 2(b) shows the resulting spectrum. The cavity spectrum now exhibits a peak at the cavity resonance, resulting in 16 times enhancement of the cavity reflected intensity compared with the bare cavity spectrum.

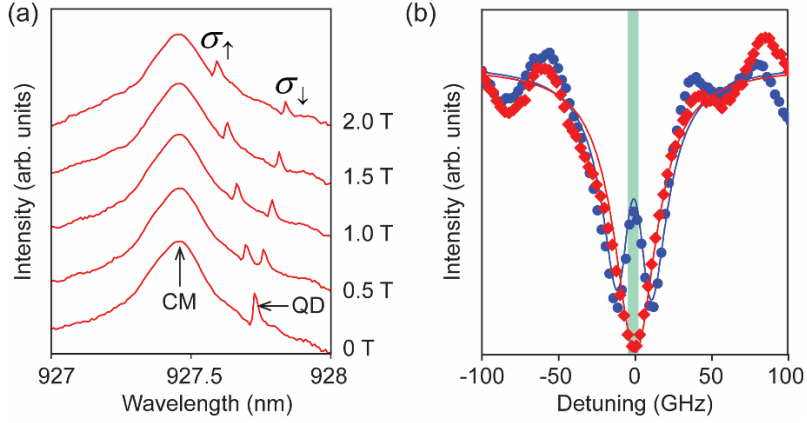

FIG. 2. (a) Cross-polarized reflectivity of the device at several different magnetic fields. (b) Cavity reflectivity at a magnetic field of 0 T (red diamonds) and 3.7 T (blue circles). Blue and red solid lines show the calculated spectra.

The solid lines in Fig. 2(b) are the calculated reflection spectra which we attain from a numerical fit to a master equation that accounts for dissipation and dephasing. We provided the details of these calculations in a previous work[41]. From the numerical fit we can extract all the parameters of the system: $g/2\pi = 10.2 \pm 0.1\,\mathrm{GHz}$, $\kappa/2\pi = 33.5 \pm 0.6\,\mathrm{GHz}$, $\Gamma_d/2\pi = 4.2 \pm 0.2\,\mathrm{GHz}$, and $\alpha = 0.92 \pm 0.01$. Using these values, we obtain a cooperativity of $C = 1.46 \pm 0.08$. We also estimate the enhancement factor to be $N/N' = 62$ using the previously reported value of $\Gamma/2\pi = 0.1\,\mathrm{GHz}$ for a bulk quantum dot[42]. The coupling strength satisfies the condition $g > \kappa/4$, indicating that we are operating at the onset of the strong coupling regime.

## IV. Measurement of cavity-enhanced spin readout

To perform spin readout, we use a pump-probe pulse sequence shown in Fig. 3(a), which measures the time evolution of the cavity reflection. We generate the pump and probe pulses out of two narrow-band continuous wave lasers, each of which is modulated by an electro-optic modulator. The pump pulse prepares the spin to either spin-up or spin-down state by resonantly pumping either the $\sigma_\downarrow$ or $\sigma_\uparrow$ transition. The probe pulse is always resonant with the cavity. We set the peak power of the pump pulse to be 710 nW, which is well beyond the saturation power for both transition $\sigma_\downarrow$ and $\sigma_\uparrow$. We set the peak power of the probe pulse to be 50 nW (measured before the objective lens), corresponding to 0.14 photons per modified lifetime of transition $\sigma_\uparrow$ (see Supplementary Material Section 3.2[35] for characterization of in-coupling efficiency from the objective lens to the cavity which is determined to be 4.5%). This probe power achieves the optimal spin readout performance (see Supplementary Material Section 4[35] for power-dependent spin readout measurement), because it is small enough to satisfy the weak excitation regime, but sufficiently large so that the spin-flip rate of the dot is dominated by photon back-action rather than the intrinsic spin decay (see Supplementary Material Section 3.3[35] for intrinsic spin-flip time measurement). We set the duration of both pump and probe pulses to be 2 µs, which is long enough compared with the spin-flip time induced by both the pump (< 6 ns) and probe fields (17.6 ns).

Figure 3(b) shows the intensity of the reflected probe pulse when we initialize the spin to the spin-up (red filled circles) and spin-down states (blue open circles) respectively. Initially, the reflected intensity for the spin-up case is 9 times higher than the spin-down case, but it decays over time due to optically induced spin flips. The red and blue solid lines show numerical fit of the measured data to an exponential function (for spin-up) and a constant background (for spin-down) respectively. The intensity for the spin-up case exponentially decays with a time constant of 17.6 ns because the probe laser induces a spin-flip. The signal decays to a finite background level which is caused by the imperfect extinction of the cavity signal and an imperfect spin initialization fidelity of 0.95. We attribute the second bump around 100 ns, which is also present when we directly inject the laser onto the detectors, to after-pulse of the photon detector. The dark counts of the detector are more than three orders of magnitude lower than the reflected intensity at the spin-down case, and therefore constitute a negligible contribution to the overall signal.

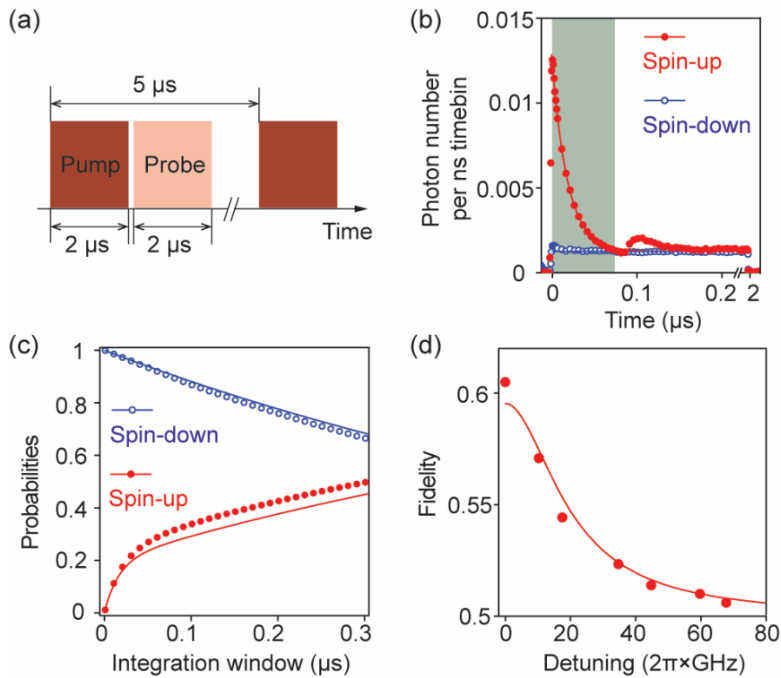

FIG. 3. (a) Pump-probe pulse sequence for spin readout measurements. (b) Intensity of the reflected probe pulse. The blue open circles and the red filled circles show the measured data when the spin is initialized in the spin-down and spin-up state respectively. The blue and red solid lines show a numerical fit to a constant and an exponential function respectively. (c) Detection probability of spin-up state (red) and spin-down state (blue) as a function of the integration window length. The blue open circles and red filled circles show the measured probability, and the blue and red solid lines show the numerical calculated probability. (d) Fidelity as a function of detuning between transition $\sigma_\uparrow$ and the cavity. Red circles show measured value at several detuning conditions, and red solid line shows a numerical fit.

The choice of integration time plays a crucial role in the spin readout scheme. Longer integration times result in a larger number of collected photons. However, as shown in Fig. 3(b), after 80 ns the spin-up state decays to the spin-down state due to optically induced spin flips. Integrating beyond this time window will only add background photons without increasing the signal.

We define $P_\uparrow$ as the probability of detecting at least one photon reflected from the cavity when the dot is initially in the spin-up state, and $P_\downarrow$ as the probability of detecting zero photons when the spin is initially in the spin-down state. Figure 3(c) plots the measured $P_\uparrow$ (red filled circles) and $P_\downarrow$ (blue open circles) as a function of the integration time. To measure these values, we repeat the pulse sequence shown in Fig. 3(a) for $n = 2{,}000{,}000$ times, and calculate the probabilities as $P_\uparrow = n_\uparrow/n$ and $P_\downarrow = 1 - n_\downarrow/n$, where $n_\uparrow$ and $n_\downarrow$ are the number of measurements that register at least a photon within an integration time window for the spin-up and spin-down initialization respectively. The red and blue solid lines show numerically calculated values for $P_\uparrow$ and $P_\downarrow$ based on the average photon number detected within each integration window obtained in Fig. 3(b). These calculations assume Poisson counting statistics for the detected photons. The small deviation between the experiment and calculation is due to detector dead time, which results in experimental counting statistics that slightly deviate from a Poisson distribution.

The probability $P_\uparrow$ initially increases rapidly as we collect more signal, but tapers off after approximately 80 ns due to collected background photons. In contrast, $P_\downarrow$ continually decreases as we increase the integration window due background photons. From these two probabilities, we can calculate the spin readout fidelity given by $F = (P_\uparrow + P_\downarrow)/2$ [10], which achieves an optimal value of $F = 0.61 \pm 0.0005$ at a window of 75 ns (indicated as the grey bar in Fig. 3(b)). At this optimal window, we detect an average number of 0.3 photons for the spin-up state, and 0.1 photons for the spin-down state. We note that because the optimal measurement time of 75 ns is longer than the laser induced spin-flip time of 17.6 ns, the measurement destroys the quantum state of the spin.

Figure 3(d) shows the measured optimal spin readout fidelity as a function of detuning $\Delta$ between transition $\sigma_\uparrow$ and the cavity. To control the detuning, we reduce the applied magnetic field, and adjust the probe center frequency to always be resonant with transition $\sigma_\uparrow$. We also optimize the detection polarization at each detuning by adding another rotatable quarter-wave plate, so that the reflected probe intensity is always maximally suppressed when the dot is in the spin-down state (see Supplementary Materials Section 5[35]). The fidelity achieves the maximum at the resonance condition, and rapidly decays as we detune from the cavity resonance, demonstrating that the improved signal is due to cavity enhancement. The red solid line in Fig. 3(d) shows a numerical calculation of the fidelity as a function of detuning $\Delta$ assuming a linear photon detector (i.e. no dead time; see Supplementary Material Section 6[35]), which agrees well with the measured results.

## V. Discussions

It is instructive to compare the performance of the cavity readout approach to what we would attain using resonance fluorescence from the bare quantum dot (not coupled to the cavity). In a resonance fluorescence measurement, the average number of photons that the bare quantum dot could emit via the cycling transition is given by $N' = \Gamma/\gamma$, or equivalently, $N' = (1-R)/R$ where $R$ is the branching ratio. In the shot noise limit, the probability $P_\uparrow$ is given by $P_\uparrow = 1 - \exp(-\eta N')$, where $\eta$ is the overall detection efficiency of emitted photons. In the absence of dark counts and background signal, the fidelity is given by $F' = \frac{1+P_\uparrow}{2} = 1 - \frac{1}{2}e^{-\eta(1-R)/R}$. This expression puts a fundamental limit on the attainable fidelity using resonance fluorescence from the bare dot.

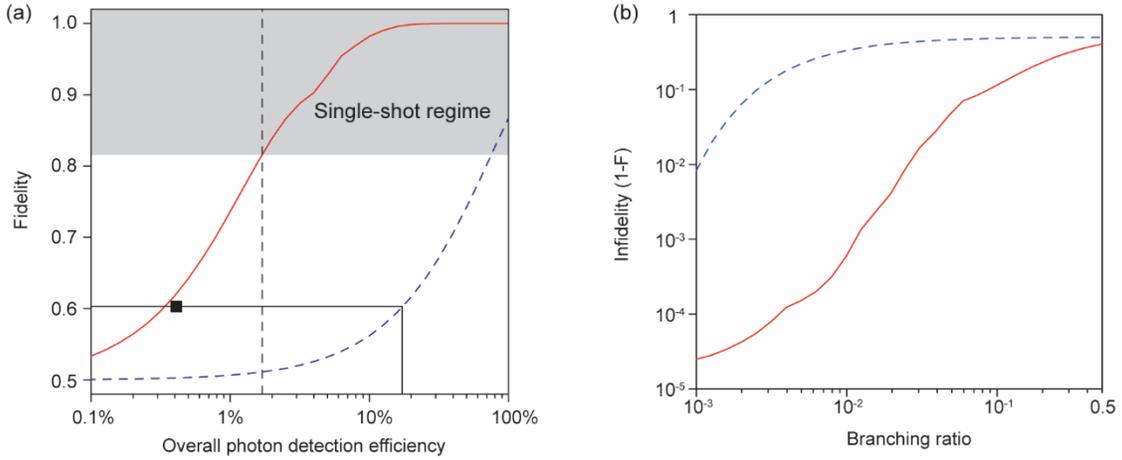

FIG. 4. (a) Expected fidelity of our system (red solid line) and the upper bound fidelity for bare dot resonance fluorescence (blue dashed line) as a function of overall photon detection efficiency. The black square shows the measured value for our current device. The vertical dashed line shows the efficiency to achieve single-shot limit using our device. (b) Expected infidelity of our system (red solid line) and the lower bound infidelity for bare dot resonance fluorescence (blue dashed line) as a function of the quantum dot branching ratio.

Figure 4(a) plots the expected fidelity of the system (red solid line) and the upper bound to the fidelity of the bare dot (blue dashed line) as a function of overall photon detection efficiency (see Supplementary Section 6[35] for a description of numerical calculation). The black square shows the experimentally measured value for using the cavity approach, which shows a spin readout fidelity of 0.61 at an overall detection efficiency of $\eta = 0.41\%$. We attribute the slight mismatch between the experimental data point and the theoretical curve to uncertainty in the efficiency, which depends on the specific alignment condition. If we read out the spin based on the resonance fluorescence from a bare dot, at the same overall photon detection efficiency, we would obtain a fidelity of $F' = 0.503$, which is very close to a fidelity of 0.5 where the measurement provides no information about the spin-state. Even with an overall photon detection efficiency of 1, the upper bound fidelity for using resonance fluorescence from a bare dot is only 0.87. This poor fidelity is due to poor branching ratio of this dot of 0.43 (see Supplementary Material Section 3.2[35] for measurement of the quantum dot branching ratio). To achieve a fidelity of $F = 0.61$ using resonance fluorescence of the bare dot would require an efficiency of 17.8%, which corresponds to 43 times improvement.

The shaded area is the region where the fidelity exceeds 0.82, which is conventionally defined as the single-shot measurement regime[10]. With the cavity quantum electrodynamics approach, single-shot readout requires an overall detection efficiency of 1.7%, which is only a factor of 4 larger than our current system efficiency. Thus, even for this dot that has a very poor branching ratio, our cavity approach is close to the single-shot regime.

We note that previous works reported a spin-readout fidelity of 0.82 using resonance fluorescence spectroscopy, which is within the single-shot limit[10]. These measurements achieved the single-shot regime because they used a quantum dot with a branching ratio of 0.002, which is more than two orders of magnitude better than the branching ratio of the dot used in this work. Figure 4(b) plots the expected infidelity $D = 1 - F$ of the cavity-enhanced readout approach (red solid line), along with the fundamental bound for the resonance fluorescence approach on a bare dot (blue dashed line) as a function of the quantum dot branching ratio. We assume an overall photon detection efficiency of 0.41% in the calculation, equal to the efficiency of our system. The results show that a branching ratio of $10^{-2}$ to $10^{-3}$, which are attainable in charge-stabilized quantum dots[10,15], would enable a readout infidelity of $10^{-3}$ to $10^{-4}$. These values are highly promising for efficient quantum error correction[43]. In contrast, bare resonance fluorescence can only attain an infidelity of only $5 \times 10^{-2}$.

## VI. Conclusions

In summary, we have demonstrated optical readout of a single solid-state spin by using strong light-matter interactions with an optical cavity. We showed that the cavity enables spin readout with a fidelity of 0.61 for a quantum dot that has a poor branching ratio of 0.43. To achieve the same value using resonance fluorescence require a factor of 43 improvement in photon collection efficiency. Our current experiment is only a factor of 4 away in efficiency from the single-shot regime. We could potentially improve this efficiency by replacing our avalanche photodiode detectors with superconducting nanowire detectors that could provide a factor of 3 increase in detection efficiency. Our collection efficiency is also low (4.5%) due to the finite numerical aperture of the objective lens. Directional photonic crystal cavity designs[44] or micro-post cavities[45,46] that provide a highly collimated transverse mode could significantly improve this efficiency. Directly extracting light to a waveguide could also increase the efficiency[47]. For quantum dot spin readout, charge-stabilized dots embedded in a diode membrane could significantly improve readout fidelity since they possess much better branching ratio. Such diode structures can also be incorporated in photonic crystal cavities as demonstrated by recent works[48,49]. Due to the scalable nature of the photonic crystal platform, our results can be directly applied to integrated devices comprised of waveguides and cavities that exhibit similar strong light-matter interactions[50]. Combining with recent developed technologies for on-chip photon detection[51,52], our results could eventually lead to chip-integrated solid-state qubit measurements, which paves the way towards quantum information processing with compact on-chip devices.

**Acknowledgements**

The authors would like to acknowledge fruitful discussions with Dr. Edward Flagg. This work is supported by the DARPA QUINESS program (grant number W31P4Q1410003), the Physics Frontier Center at the Joint Quantum Institute, the National Science Foundation (grant number PHYS. 1415458), and the Center for Distributed Quantum Information.

# Supplementary Materials

# Cavity-enhanced optical readout of a single solid-state spin

Shuo Sun, Hyochul Kim, Glenn S. Solomon, and Edo Waks

## 1. Derivation of collection probability

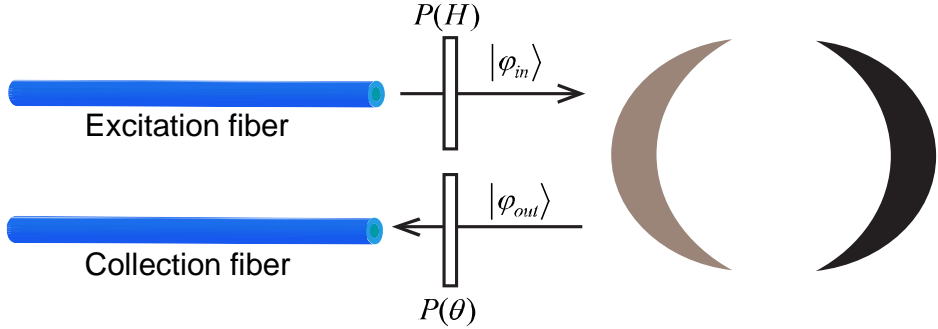

**Supplementary Figure 1**. Schematic of the experimental setup in the theoretical model.

Supplementary Figure 1 illustrates the model we use to describe the system. We calculate the collection probability of a cavity-coupled incident photon after its reflection. We first provide an intuitive picture by considering the ideal limit where the transverse spatial mode of the cavity matches perfectly with the excitation and collection fibers. In this limit, we only need to focus on the polarization degree of freedom of the photon. We express the state of the incident photon as $|\psi\rangle_{in} = \frac{1}{\sqrt{2}}(|x\rangle + |y\rangle)$, where $|x\rangle$ and $|y\rangle$ represents the polarization that is perpendicular and parallel with the cavity mode respectively. Since only the $y$-polarized photon couples to the cavity, the reflected field is given by $|\psi\rangle_{out} = \frac{1}{\sqrt{2}}(|x\rangle + r_{\uparrow,\downarrow}|y\rangle)$. We set the angle of the detection polarizer to be $\theta = 45°$, corresponding to detection along the same polarization as the

incident field. The collection probability of the photon is thus given by $P = \frac{1}{4}|1+r_{\uparrow,\downarrow}|^2$.
Therefore, the detector will not detect any photons when the spin is in the spin-down state ($r_{\downarrow} = -1$), but will detect a bright photon flux for spin-up state ($r_{\uparrow} = 1$).

When the spatial mode of the cavity does not perfectly match the excitation and collection fibers, the collection fiber may collect spurious reflection from the surface in addition to the component that is coupled to the cavity. We now show that by simply changing $\theta$, we can compensate for this mode mismatch. Specifically, by properly selecting an optimal $\theta$ we attain a collection probability of $P' = \frac{1}{4}\beta\beta'|1+r_{\uparrow,\downarrow}|^2$ for an incident photon injected from the excitation fiber, where $\beta'$ is the coupling efficiency from the excitation fiber into the cavity, and $\beta$ is the coupling efficiency from the cavity to the collection fiber. The collection probability $P$ for a cavity-coupled photon is therefore given by $P = P'/\beta' = \frac{1}{4}\beta|1+r_{\uparrow,\downarrow}|^2$.

We express the wavefunction of a photon in the basis $|M, Pol\rangle$, where $Pol$ and $M$ represents the polarization and transverse spatial mode of the photon respectively. The wavefunction of an incident photon $|\varphi_{in}\rangle$ in the excitation fiber is given by

$$|\varphi_{in}\rangle = |f_1, H\rangle, \tag{1}$$

where $|H\rangle$ is the polarization state of the incident photon set by the excitation polarizer, satisfying $|H\rangle = (|x\rangle + |y\rangle)/\sqrt{2}$, and $|f_1\rangle$ is the transverse spatial mode of the excitation fiber. We can express the transverse spatial mode of the photon as a superposition of a perfectly mode matched component represented by $|a\rangle$ and an orthogonal component $|b\rangle$ that does not match the cavity mode using Schmidt decomposition, given by $|f_1\rangle = \sqrt{\beta'}|a\rangle + \sqrt{1-\beta'}|b\rangle$. Under this

transformation, we can rewrite the wavefunction of the incident photon as

$$|\varphi_{in}\rangle = \sqrt{\frac{\beta'}{2}}(|a,x\rangle + |a,y\rangle) + \sqrt{\frac{1-\beta'}{2}}(|b,x\rangle + |b,y\rangle). \tag{2}$$

The state of the reflected photon $|\varphi_{out}\rangle$ is therefore given by

$$|\varphi_{out}\rangle = \sqrt{\frac{\beta'}{2}}(|a,x\rangle + r_{\uparrow(\downarrow)}|a,y\rangle) + \sqrt{\frac{1-\beta'}{2}}(|b,x\rangle + |b,y\rangle), \tag{3}$$

where only the mode component $|a,y\rangle$ picks up a spin-dependent phase, all other components are directly reflected because they do not couple to the cavity. The reflected photon goes through a detection polarizer whose polarization axis is set at $|\theta\rangle = |x\rangle\cos\theta + |y\rangle\sin\theta$, and then collected by the collection fiber with a transverse spatial mode $|f_2\rangle$. We can therefore calculate the collection probability $P'$ as

$$P' = |\langle f_2, \theta | \varphi_{out}\rangle|^2 = \left|\sqrt{\frac{\beta\beta'}{2}}(\cos\theta + r_{\uparrow(\downarrow)}\sin\theta) + \sqrt{\frac{(1-\beta)(1-\beta')}{2}}(\cos\theta + \sin\theta)\right|^2. \tag{4}$$

One can easily verify that if we choose $\theta$ equal to

$$\theta = \arctan\left(-\frac{\sqrt{(1-\beta)(1-\beta')} + \sqrt{\beta\beta'}}{\sqrt{(1-\beta)(1-\beta')} - \sqrt{\beta\beta'}}\right), \tag{5}$$

the probability $P'$ is given by

$$P' = \frac{1}{2}\beta\beta'\sin^2\theta\left|1 + r_{\uparrow(\downarrow)}\right|^2. \tag{6}$$

In practice, we have $\beta, \beta' \ll 1$ because our cavity has a largely divergent mode that is not well matched to a fiber. Under this approximation, we have $\theta \approx \pi/4$. We can therefore write the probability $P'$ as

$$P' \cong \frac{1}{4}\beta\beta'\left|1+r_{\uparrow(\downarrow)}\right|^2. \tag{7}$$

which has exactly the same expression as the perfect mode-matching case, except for an additional linear coefficient $\beta\beta'$ that accounts for coupling losses due to mode mismatch. The collection probability $P$ for a cavity-coupled photon is therefore given by $P = P'/\beta' = \frac{1}{4}\beta\left|1+r_{\uparrow,\downarrow}\right|^2$.

## 2. Derivation of cavity enhanced spin readout signal

In this Section, we calculate the average number of photons reflected into the detection polarization basis before the incident field induces a spin-flip. To derive an analytical solution, we only consider the coupling between transition $\sigma_\uparrow$ and the cavity, and ignore the coupling between transition $\sigma_\downarrow$ and the cavity due to its large detuning. Under this assumption, we could use a simplified three-level energy structure to model the quantum dot, as shown in Supplementary Figure 2. The two ground states, labeled as $|1\rangle$ and $|2\rangle$ in the figure, represent the quantum dot spin ground states $|\uparrow\rangle$ and $|\downarrow\rangle$ respectively. The excited state, labeled as $|3\rangle$ in the figure, represents the trion state $|\Uparrow\rangle$. In order to obtain an intuitive analytical expression for the photon number, we also assume that the emitter dephasing and spectral diffusion are absent, and that the cavity couples only to its reflection mode. In addition, we focus our calculation in the ideal case where the transverse spatial mode of the incident field matches perfectly with the cavity. As discussed in Section 3, under this assumption, the detection polarizer is set at the *H* polarization direction. The obtained results can be easily generalized to imperfect mode-matching conditions using the method described in Section 1.

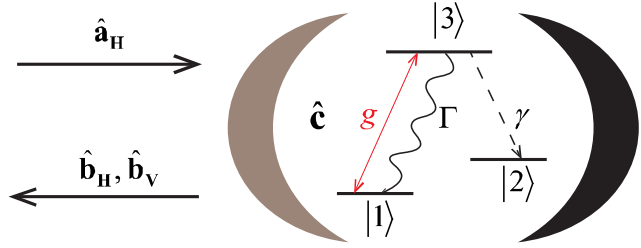

**Supplementary Figure 2**. Theoretical model for calculation of cavity enhanced spin readout signal.

We denote $\hat{\mathbf{a}}_\mathrm{H}$ as the bosonic annihilation operator for the incident field that is polarized in the *H* direction, $\hat{\mathbf{b}}_\mathrm{H}$ and $\hat{\mathbf{b}}_\mathrm{V}$ as the bosonic annihilation operator for the reflected field that is polarized in *H* and *V* directions respectively, and $\hat{\mathbf{c}}$ as the bosonic annihilation operator for the cavity field. We denote *N* as the average number of photons reflected into the *H* polarization basis before the probe flips the quantum dot from spin-up to spin-down state. We calculate *N* as $N = P_{ref}/P_{flip}$, where $P_{ref}$ is the probability that a single photon is reflected into the *H* polarization mode when the quantum dot is initially in state $|1\rangle$, and $P_{flip}$ is the probability that a single photon flips the spin from state $|1\rangle$ to state $|2\rangle$. The expression for $P_{ref}$ is given by

$$P_{ref} = \int_{-\infty}^{\infty} \hat{\mathbf{b}}_\mathrm{H}^\dagger(t)\hat{\mathbf{b}}_\mathrm{H}(t)\,dt, \tag{8}$$

where the output operator $\hat{\mathbf{b}}_\mathrm{H}$ can be calculated from the input operator $\hat{\mathbf{a}}_\mathrm{H}$ using the cavity input-output formalism[1], given by

$$\hat{\mathbf{b}}_{\mathbf{H}}(t) = \hat{\mathbf{a}}_{\mathbf{H}}(t) - \sqrt{\frac{\kappa}{2}}\hat{\mathbf{c}}(t). \tag{9}$$

The expression for $P_{flip}$ is given by

$$P_{flip} = \gamma \int_{-\infty}^{\infty} \hat{\boldsymbol{\sigma}}_{33}(t) dt = \gamma \int_{-\infty}^{\infty} \hat{\boldsymbol{\sigma}}_{13}^{\dagger}(t) \hat{\boldsymbol{\sigma}}_{13}(t) dt, \tag{10}$$

where we denote $\hat{\boldsymbol{\sigma}}_{ij} = |i\rangle\langle j|$ ($i, j \in \{1,2,3\}$), and $\gamma$ is the decay rate from the excited state $|3\rangle$ to ground state $|2\rangle$.

In order to calculate $P_{ref}$ and $P_{flip}$, we need an expression for the cavity field operator $\hat{\mathbf{c}}(t)$ and the atomic dipole operator $\hat{\boldsymbol{\sigma}}_{13}(t)$. We derive these expressions using the Heisenberg-Langevin formalism. The Hamiltonian for the coupled cavity and quantum dot system in the rotating reference frame with respect to the incident field frequency $\omega$ is given by

$$\hat{\mathbf{H}} = \hbar(\omega_c - \omega)\hat{\mathbf{c}}^{\dagger}\hat{\mathbf{c}} + \hbar(\omega_x - \omega)\hat{\boldsymbol{\sigma}}_{33} + i\hbar g\left(\hat{\mathbf{c}}\hat{\boldsymbol{\sigma}}_{31} - \hat{\boldsymbol{\sigma}}_{13}\hat{\mathbf{c}}^{\dagger}\right), \tag{11}$$

where $\omega_c$ and $\omega_x$ are the resonance frequencies of the cavity mode and quantum dot transition $|1\rangle \leftrightarrow |3\rangle$ respectively, and $g$ is the coupling strength between the cavity mode and quantum dot transition $|1\rangle \leftrightarrow |3\rangle$. In the weak field limit, the Heisenberg-Langevin equations are given by[2]

$$\frac{d\hat{\mathbf{c}}}{dt} = -\left[i(\omega_c - \omega) + \frac{\kappa}{2}\right]\hat{\mathbf{c}} - g\hat{\boldsymbol{\sigma}}_{13} + \sqrt{\frac{\kappa}{2}}\hat{\mathbf{a}}_{\mathbf{H}}, \tag{12}$$

$$\frac{d\hat{\boldsymbol{\sigma}}_{13}}{dt} = -\left[i(\omega_x - \omega) + \frac{\Gamma + \gamma}{2}\right]\hat{\boldsymbol{\sigma}}_{13} + g\hat{\mathbf{c}}, \tag{13}$$

where $\kappa$ is the cavity energy decay rate, $\Gamma$ is the decay rate from the excited state $|3\rangle$ to ground states $|1\rangle$. We calculate the cavity field operator $\hat{\mathbf{c}}$ and the atomic dipole operator $\hat{\boldsymbol{\sigma}}_{13}$ by taking the steady solution of Eqs. (12) and (13), given by

$$\hat{c} = \frac{\sqrt{\frac{\kappa}{2}}\left[i(\omega_x - \omega) + \frac{\Gamma + \gamma}{2}\right]}{\left[i(\omega_c - \omega) + \frac{\kappa}{2}\right] \cdot \left[i(\omega_x - \omega) + \frac{\Gamma + \gamma}{2}\right] + g^2} \hat{a}_H, \quad (14)$$

$$\hat{\sigma}_{13} = \frac{g\sqrt{\frac{\kappa}{2}}}{\left[i(\omega_c - \omega) + \frac{\kappa}{2}\right] \cdot \left[i(\omega_x - \omega) + \frac{\Gamma + \gamma}{2}\right] + g^2} \hat{a}_H. \quad (15)$$

We obtain $P_{ref}$ and $P_{flip}$ by substituting Eqs. (14) and (15) into Eqs. (8) - (10). At the resonance condition where $\omega_c = \omega_x = \omega$, the expressions for $P_{ref}$ and $P_{flip}$ are given by

$$P_{ref} = \frac{g^4}{\left(\frac{\kappa}{2} \cdot \frac{\Gamma + \gamma}{2} + g^2\right)^2}, \quad (16)$$

$$P_{flip} = \frac{g^2 \kappa \gamma}{2\left(\frac{\kappa}{2} \cdot \frac{\Gamma + \gamma}{2} + g^2\right)^2}. \quad (17)$$

We therefore calculate $N$ as $N = P_{ref}/P_{flip} = 2g^2/\kappa\gamma$.

## 3. Device characterization

### 3.1 Verification of a negatively charged quantum dot

To verify that the measured quantum dot is charged, we measure the photoluminescence of the quantum dot with a magnetic field applied perpendicular to the quantum dot growth direction (Voigt configuration). Supplementary Figure 3 shows the photoluminescence of the quantum dot as we vary the magnetic field amplitude. The quantum dot emission (labeled as "QD") splits into four peaks (labeled as "1" to "4") as we increase the magnetic field, demonstrating that this dot is charged. In this measurement, we have redshifted the cavity resonance far away from the quantum dot by ~ 1 nm by deposition of nitrogen gas[3].

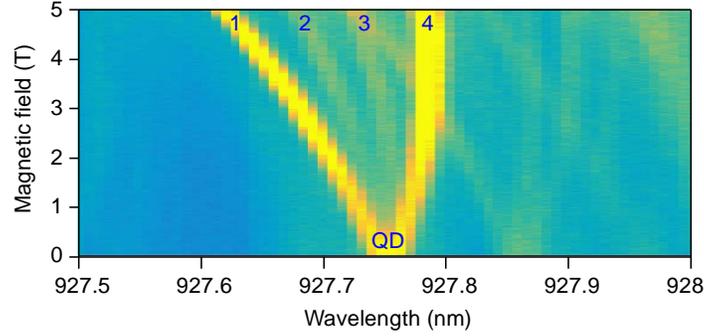

**Supplementary Figure 3**. Photoluminescence of the quantum dot as we vary the amplitude of the magnetic field applied in the Voigt geometry.

To determine whether the dot is positively or negatively charged, we measure the frequency difference $\Delta_e$ between the two spin ground states using the spectrum acquired at 5 T. From this measurement, we calculate the Lande g-factor to be $g_l = 0.53$ using the relation $g_l = \hbar\Delta_e/\mu_B B$. This value is consistent with the typically reported numbers for a quantum dot containing a single electron that range from 0.4 to 0.6[4-11]. Positively charged quantum dots containing a hole spin exhibit a Lande g-factor below 0.3[12,13], much smaller than our measured values, indicating that our spin is originating from an additional electron in the quantum dot.

### 3.2 Measurement of branching ratio

The branching ratio $R_B$ of a quantum dot is given by $R_B = \gamma/(\Gamma+\gamma)$, where $\Gamma$ is the spontaneous emission rate of transition $\sigma_\uparrow$ for a bare quantum dot, and $\gamma$ is the decay rate of the optical forbidden transition $|\Uparrow\rangle \to |\downarrow\rangle$. The value of $\Gamma$ is consistent among different dots, therefore we obtain that $\Gamma/2\pi = 0.1\,\text{GHz}$ from previous studies on bare InAs quantum dots[14]. The value of $\gamma$ can vary significantly among dots, so we must measure it independently for the one we reported in the manuscript. Since $\gamma$ represents a nonradiative decay rate, its value is not affected by the cavity and can be measured experimentally.

We perform the same measurement as described in Fig. 3(a) of the main manuscript, but vary the power of the probe pulse. Supplementary Figure 4(a) shows the time resolved reflected intensity for several different probe pulse powers (measured before objective lens). The signal for all probe powers exponentially decays, which is due to optically induced spin-flip as discussed in Fig. 3(b) of the main text. The spin-flip rate $\gamma_p$ increases when we increase the power of the probe pulse because the probe pulse drives transition $\sigma_\uparrow$ more efficiently. The red circles in Supplementary Figure 4(b) shows the extracted spin-flip rate as a function of probe power. When the probe is strong enough to saturate the strongly coupled system, $\gamma_p$ no longer increases and saturates at a value that is determined by $\gamma$. We thus obtain $\gamma$ by fitting the measured data to a theoretical model as described below (red solid line).

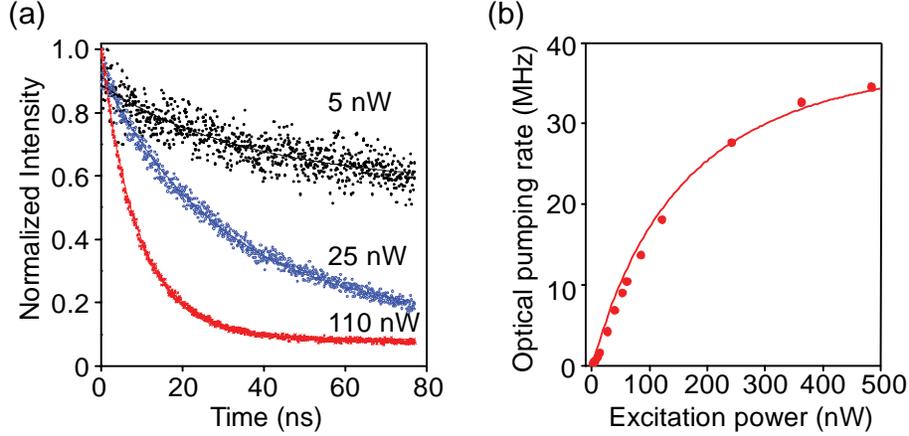

**Supplementary Figure 4**. (a) Time resolved reflected intensity at several different probe powers. (b) Optical pumping rate as a function of probe power. The excitation power is measured before the objective lens. Red circles show the measured data and red solid line shows numerically calculated values.

We again model the quantum dot using a simplified three-level energy structure as illustrated in Supplementary Figure 2. In our model, only transition $|1\rangle \leftrightarrow |3\rangle$ couples to the cavity, with a coupling strength $g$. An external laser with frequency $\omega$ drives the cavity at $t > 0$. We set the initial state of the quantum dot to be in state $|1\rangle$ at time $t = 0$, and calculate the occupation probability of state $|3\rangle$ at an arbitrary time $t$, denoted by $P_3(t)$, given by $P_3(t) = Tr(\rho(t)\hat{\sigma}_{33})$, where $\rho$ is the density matrix of the system, and we denote $\hat{\sigma}_{ij} = |i\rangle\langle j|$ ($i, j \in \{1, 2, 3\}$). We can thus directly obtain $\gamma_p$ from the rate that $P_3(t)$ increases, given by $\gamma_p = \frac{d}{dt}P_3(t) \big/ P_3(t)$. In practice, we choose to calculate $\gamma_p$ at large enough $t$ to get rid of the transient response of the system.

We obtain the density matrix $\rho(t)$ by numerically solving the master equation of the system, given by $d\rho/dt = -i/\hbar\left[\hat{\mathbf{H}}, \rho\right] + \hat{\mathbf{L}}\rho$, where $\hat{\mathbf{H}}$ is the system Hamiltonian, and $\hat{\mathbf{L}}$ is the Liouvillian superoperator that accounts for non-unitary evolution due to all dissipative mechanisms. The Hamiltonian, expressed in the reference frame rotating with respect to the input laser frequency $\omega$, is given by

$$\hat{\mathbf{H}} = \hbar\Delta_c \hat{\mathbf{c}}^\dagger \hat{\mathbf{c}} + \hbar\Delta_x \hat{\boldsymbol{\sigma}}_{33} + i\hbar g\left(\hat{\mathbf{c}}\hat{\boldsymbol{\sigma}}_{31} - \hat{\boldsymbol{\sigma}}_{13}\hat{\mathbf{c}}^\dagger\right) + i\hbar\sqrt{\frac{\kappa_{ex}}{2}}\varepsilon\left(\hat{\mathbf{c}}^\dagger - \hat{\mathbf{c}}\right), \tag{18}$$

where $\hat{\mathbf{c}}$ is the photon annihilation operator for the cavity mode, $\Delta_c$ and $\Delta_x$ are given by $\Delta_c = \omega_c - \omega$ and $\Delta_x = \omega_x - \omega$ respectively, where $\omega_c$ and $\omega_x$ are the resonant frequencies of the cavity mode and transition $|1\rangle \leftrightarrow |3\rangle$ respectively, $\kappa_{ex}$ is the cavity energy decay rate to its reflection mode, given by $\kappa_{ex} = \alpha\kappa$, where $\alpha$ is the interference contrast and $\kappa$ is the cavity total energy decay rate, $\varepsilon$ is the amplitude of the probe field, given by $\varepsilon = \sqrt{\frac{\beta' P}{\hbar\omega}}$, where $\beta'$ is the coupling efficiency from the objective lens to the cavity, and $P$ is the power of the incident field measured before the objective. The Liouvillian superoperator $\hat{\mathbf{L}}$ is given by

$$\hat{\mathbf{L}} = \kappa D(\hat{\mathbf{a}}) + \Gamma D(\hat{\boldsymbol{\sigma}}_{13}) + \gamma D(\hat{\boldsymbol{\sigma}}_{23}) + 2\gamma_d D(\hat{\boldsymbol{\sigma}}_{33}), \tag{19}$$

where $D(\hat{\mathbf{O}})\rho = \hat{\mathbf{O}}\rho\hat{\mathbf{O}}^\dagger - 1/2\hat{\mathbf{O}}^\dagger\hat{\mathbf{O}}\rho - 1/2\rho\hat{\mathbf{O}}^\dagger\hat{\mathbf{O}}$ is the general Linblad operator form for the collapse operator $\hat{\mathbf{O}}$, $\Gamma$ and $\gamma$ are the decay rate from the quantum dot excited state $|3\rangle$ to ground states $|1\rangle$ and $|2\rangle$ respectively, and $\gamma_d$ is the pure dephasing rate for the excited state $|3\rangle$.

We numerically calculated $\gamma_p$ for each incident laser power $P$, and fit the calculated values to the measured data. In the numerical fit, we fix all other parameters using the values their measured values, except for $\gamma$ and the in-coupling efficiency $\beta'$. These fixed parameters are given by $\Delta_c = \Delta_x = 0$, $g/2\pi = 10.2$ GHz, $\kappa/2\pi = 33.5$ GHz, $\Gamma/2\pi = 0.1$ GHz, $\gamma_d/2\pi = 4.2$ GHz, and $\alpha = 0.92$. From the fit we obtain that $\gamma/2\pi = 0.075 \pm 0.002$ GHz, and therefore the branching ratio is given by $R_B = \gamma/(\Gamma + \gamma) = 0.43$.

The branching ratio of our quantum dot is much poorer than previous reported values based on charge tunable quantum dots embedded in a Schottky diode[15,16]. We attribute the large branching ratio of our dot to unstable confinement of the extra electron. Upon resonance excitation, the extra electron confined in the dot has some probability to escape from the quantum dot via an Auger process[17-20]. When the quantum dot re-traps the electron, the electron no longer possesses its previous spin state, which creates an effective spin flip channel upon resonance excitation. Using a stably charged quantum dot (e.g. ones embedded in a Schottky diode) would resolve this issue and result in a much lower branching ratio.

From the fit, we also obtain the in-coupling efficiency to be $\beta' = 4.5\%$. Since we use the same type of fiber for both excitation and collection, we estimate the collection efficiency is the same as the in-coupling efficiency. We therefore have $\beta = \beta' = 4.5\%$.

### 3.3 Measurement of intrinsic spin-flip time ($T_1$ time)

We measure the intrinsic spin flip time ($T_1$ time) through a pump-probe pulse sequence. We use the same pulse sequence as shown in Fig. 3(a) of the main text. Here we fix the frequency of the pump pulse to be resonant with the $\sigma_\uparrow$ transition, which prepares the spin in the spin-down state. We set the pulse repetition time to be 10 μs.

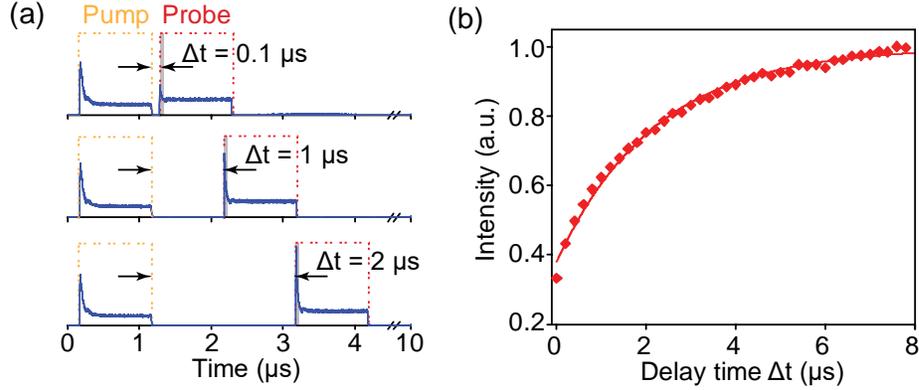

**Supplementary Figure 5**. (a) Pulse sequence for spin relaxation time measurements. The dashed lines show the illustration of the pump-probe sequence, and the blue solid lines show the histogram of the collected photons in 10 seconds. (b) Normalized intensity of the probe signal as a function of pump-probe delay time. Red diamonds and red solid line shows the measured data and the numerical fit respectively.

The blue solid lines in Supplementary Figure 5(a) show the intensity of the reflected pump and probe pulse measured at several different pump-probe delay $\Delta t$. For a short pump-probe delay, the spin mainly remains in the spin-down state when the probe excites the cavity, therefore peak intensity is very close to the background level. As we increase the pump-probe delay, the spin has a larger probability to flip to the spin-up state before the probe, which results in an increase in the peak intensity.

Supplementary Figure 5(b) shows the measured intensity of the reflected probe pulse within its first 50 ns (corresponding to the grey bars in Supplementary Figure 5(a)) as a function of the pump-probe delay. We numerically fit the measured data (red diamonds) to an exponential function (red solid line), from which we calculate the $T_1$ time to be $T_1 = 2.21 \pm 0.15\ \mu s$. The spin flip time is 2 orders of magnitude smaller than previously reported values with a quantum dot embedded in a Schottky diode[21,22] as a result of the unstable charging of our dot.

## 4. Power-dependent spin readout measurements

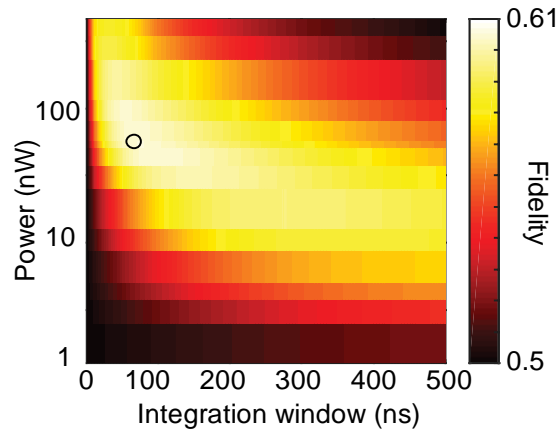

**Supplementary Figure 6**. Experimentally measured spin readout fidelity as a function of probe laser power (measured before the objective lens) and the length of the integration window. The black circle indicates the condition to achieve the optimal readout fidelity for our device.

To achieve the optimal spin readout performance, it is essential that the probe power is small enough to satisfy the weak excitation regime, but yet sufficiently large so that the spin-flip rate of the dot is dominated by photon back-action rather than the intrinsic spin decay. To verify such conditions experimentally, we repeat the same pump-probe sequence as shown in Fig. 3(a) of the main text, but vary the probe power. Supplementary Figure 6 shows the measured spin readout fidelity as a function of both probe power and integration time window. At small powers, the required integration window to achieve the optimal fidelity decreases as we increase the probe power, due to a faster measurement induced back-action. The optimal spin readout fidelity increases as we increase the probe power due to the better signal to background ratio. However, at large powers, the optimal integration window does not decrease further, and the optimal fidelity starts to degrade. This is because the probe power is large enough to saturate the emitter and starts to violate the weak excitation regime. The black circle indicates operation condition to achieve the best spin readout fidelity of 0.61. This is the probe power we used for all the reported results shown in the main text.

## 5. Detection polarization for spin readout at detuned conditions

We first provide an intuitive understanding on why we need to re-optimize the detection polarization when we detune transition $\sigma_\uparrow$ from the cavity. We provide an example in the ideal case where the transverse spatial mode of the incident field matches perfectly with the cavity. But the obtained results can be easily generalized to the imperfect mode-matching conditions using the method described in Section 1.

Supplementary Figure 7(a) and 7(b) shows calculated cavity reflection spectrum when transition $\sigma_\uparrow$ is resonant with the cavity and detuned from the cavity by 10 GHz respectively. In the calculation, we use all parameters based on the reported device, and set the polarization of both excitation and detection as *H*, same as the polarization configuration in our spin readout measurements. The blue lines show the spectra when the quantum dot is in spin-down so that the cavity is uncoupled, and the red lines show the spectra when the quantum dot is in spin-up. The vertical dashed lines in these two figures indicate the frequency of the probe laser for spin readout, which is always resonant with transition $\sigma_\uparrow$ to achieve the maximum contrast. The contrast in cavity reflectivity between spin-up and spin-down degrades significantly as transition $\sigma_\uparrow$ is detuned from the cavity, because the reflectivity of a bare cavity achieves minimum at the cavity resonance but becomes much larger at detuned conditions.

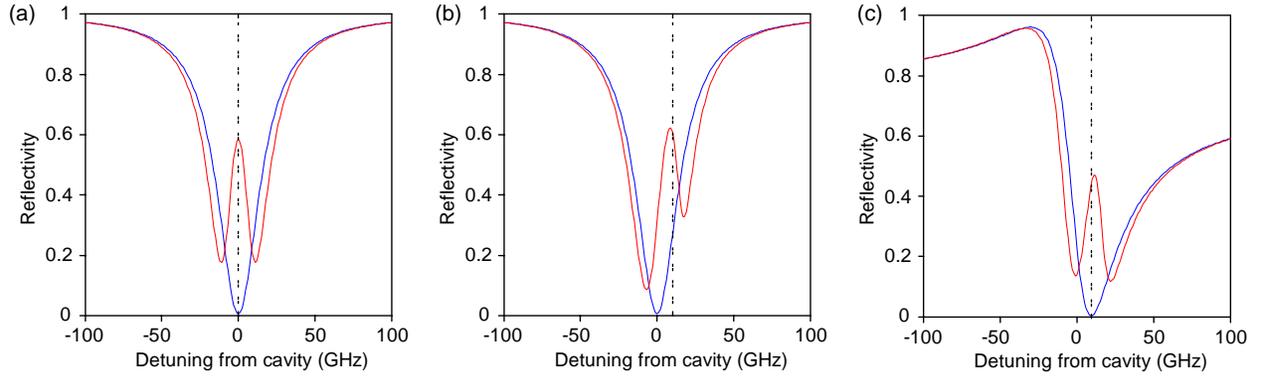

**Supplementary Figure 7** (a) Calculated cavity reflection spectrum when transition σ↑ is resonant with the cavity and the detection polarization is *H*. (b) Calculated cavity reflection spectrum when transition σ↑ is blue detuned from the cavity by 10 GHz and the detection polarization is *H*. (c) Calculated cavity reflection spectrum when transition σ↑ is blue detuned from the cavity by 10 GHz and the detection polarization is optimized to suppress the bare cavity reflectivity at 10 GHz. In all panels, the blue and red solid line represent the cases when the quantum dot is in spin-down and spin-up states respectively.

We now show that for an arbitrary detuning $\Delta$, we can always find a polarization basis such that the bare cavity reflectivity measured at this basis achieves minimum at detuning $\Delta$. We start with a general expression for the detection polarization basis, given by

$$|\theta,\varphi\rangle = |H\rangle\cos\theta + e^{i\varphi}|V\rangle\sin\theta, \qquad (20)$$

where $\theta$ and $\varphi$ are real numbers. The cavity output field in the polarization basis $|\theta,\varphi\rangle$,

denoted using the annihilation operator $\hat{\mathbf{b}}_{\theta,\varphi}$, can be written as

$$\hat{\mathbf{b}}_{\theta,\varphi} = \hat{\mathbf{b}}_{\mathbf{H}} \cos\theta + \hat{\mathbf{b}}_{\mathbf{V}} e^{i\varphi} \sin\theta, \tag{21}$$

where $\hat{\mathbf{b}}_{\mathbf{H}}$ and $\hat{\mathbf{b}}_{\mathbf{V}}$ are annihilation operators for cavity output field in the polarization basis $|H\rangle$ and $|V\rangle$ respectively. Using cavity input-output formalism[1], we have

$$\hat{\mathbf{b}}_{\mathbf{H}} = \hat{\mathbf{a}}_{\mathbf{H}} - \sqrt{\frac{\kappa_{ex}}{2}} \hat{\mathbf{c}}, \tag{22}$$

$$\hat{\mathbf{b}}_{\mathbf{V}} = -\sqrt{\frac{\kappa_{ex}}{2}} \hat{\mathbf{c}}, \tag{23}$$

where $\hat{\mathbf{a}}_{\mathbf{H}}$ is the annihilation operator for the incident field, and $\hat{\mathbf{c}}$ is the annihilation operator for the cavity field. For a bare cavity, the cavity field operator $\hat{\mathbf{c}}$ is given by

$$\hat{\mathbf{c}} = \sqrt{\frac{\kappa_{ex}}{2}} \hat{\mathbf{a}}_{\mathbf{H}} \bigg/ \left(\frac{\kappa}{2} + i\Delta\right). \tag{24}$$

We can therefore relate the bare cavity output field operator $\hat{\mathbf{b}}_{\theta,\varphi}$ with the input field operator $\hat{\mathbf{a}}_{\mathbf{H}}$ by substituting Eqns. (22) - (24) into Eq. (21). In the ideal limit where $\kappa_{ex} = \kappa$, the bare cavity output field operator $\hat{\mathbf{b}}_{\theta,\varphi}$ takes a simple expression given by

$$\hat{\mathbf{b}}_{\theta,\varphi} = \hat{\mathbf{a}}_{\mathbf{H}} \left(\cos\theta \frac{i\Delta}{\kappa/2 + i\Delta} - e^{i\varphi} \sin\theta \frac{\kappa/2}{\kappa/2 + i\Delta}\right). \tag{25}$$

Clearly, we have $\hat{\mathbf{b}}_{\theta,\varphi} = \hat{\mathbf{0}}$ if $\theta = \arctan\left(\frac{2\Delta}{\kappa}\right)$ and $\varphi = \frac{\pi}{2}$. Therefore, if we measure cavity reflectivity at this polarization basis, we will obtain vanished reflection intensity for a bare cavity reflectivity at detuning $\Delta$. In general, we can express the cavity output field operator in this polarization basis as

$$\hat{\mathbf{b}}_\Delta = \frac{\kappa/2}{\kappa/2 + i\Delta}\hat{\mathbf{a}}_\mathbf{H} - \sqrt{\frac{\kappa_{ex}}{2}}\hat{\mathbf{c}}, \tag{26}$$

obtained by substituting the optimal values of $\theta = \arctan\left(\frac{2\Delta}{\kappa}\right)$ and $\varphi = \frac{\pi}{2}$ into Eq. (21). Here we use $\hat{\mathbf{b}}_\Delta$ to denote $\hat{\mathbf{b}}_{\theta,\varphi}$ at a specific detuning $\Delta$, and the operator $\hat{\mathbf{c}}$ obeys the Heisenberg-Langevin equations given by Eq. (12) and (13).

Supplementary Figure 7(c) shows the calculated cavity reflection spectrum obtained at the optimized polarization basis for $\Delta = 10 \text{ GHz}$. We again calculate the cavity reflectivity for both spin-up and spin-down cases, shown as the blue and red solid lines. Clearly, the optimal detection polarization allows much better reflectivity contrast between spin-up and spin-down cases.

In experiments, to optimize the detection polarization for each detuning condition $\Delta$, we firstly turn off the white-light illumination that is used for stabilizing the charge in the quantum dot. Under this condition, the quantum dot becomes a neutral exciton which is far detuned from the cavity, and the cavity behaves as an empty one. We then park a continuous-wave laser at the probe frequency and minimize the reflected intensity of the laser from the bare cavity by adjusting a half-wave and quarter-wave plate placed before the detection polarizer. After that we fix the two waveplates and perform spin readout measurements.

## 6. Numerical calculation of spin readout fidelity

The general expression for fidelity is given by $F = (P_\uparrow + P_\downarrow)/2$, where $P_\uparrow$ and $P_\downarrow$ are the probabilities of getting the correct spin measurement when the quantum dot is initially in spin-up and spin-down states respectively. In the calculation for Fig. 3(d) (red solid line) of the main manuscript, we take $P_\uparrow$ as the probability of detecting at least one photon when the dot is initially in the spin-up state, and $P_\downarrow$ as the probability of detecting zero photons when the spin is initially in the spin-down state. In the shot noise limit, $P_\uparrow$ and $P_\downarrow$ are given by $P_\uparrow = \max_T \{1 - \exp(-N_\uparrow(T))\}$ and $P_\downarrow = \max_T \{1 - \exp(-N_\downarrow(T))\}$ respectively, where $T$ is the time duration of the integration window, and $N_\uparrow(T)$ and $N_\downarrow(T)$ are the average number of detected photons within an integration time $T$ when the quantum dot is initially in the spin-up and spin-down states respectively.

The expressions for $N_\uparrow(T)$ and $N_\downarrow(T)$ are given by

$$N_\uparrow(T) = \eta \int_0^T Tr\left(\rho_\uparrow(t) \hat{\mathbf{b}}_\Delta^\dagger \hat{\mathbf{b}}_\Delta\right), \tag{27}$$

$$N_\downarrow(T) = \eta \int_0^T Tr\left(\rho_\downarrow(t) \hat{\mathbf{b}}_\Delta^\dagger \hat{\mathbf{b}}_\Delta\right), \tag{28}$$

where $\eta$ is the overall photon detection efficiency, $\rho_\uparrow(t)$ and $\rho_\downarrow(t)$ are the density matrixes of the system at time $t$ when the initial states of the quantum dot are in spin-up and spin-down respectively, $\hat{\mathbf{b}}_\Delta$ is the bosonic annihilation operator for the output field in the optimized detection polarization basis for detuning $\Delta$ and is given by Eq. (26). We numerically obtain $N_\uparrow(T)$ and $N_\downarrow(T)$ by solving the system master equation $d\rho_{\uparrow(\downarrow)}/dt = -i/\hbar\left[\hat{\mathbf{H}}, \rho_{\uparrow(\downarrow)}\right] + \hat{\mathbf{L}}\rho_{\uparrow(\downarrow)}$, where the Hamiltonian and Liouvillian superoperator are given by Eq. (18) and (19) respectively.

For the calculation shown in Fig. 4 of the main text, we adapt a more generalized criterion for determining whether the quantum dot is in spin-up or spin-down. Previously we determine spin-down if we do not detect any photons, and spin-up if we detect one or more photons. While this criterion works well at small $\eta$, it is no longer valid at large $\eta$ where the background counts become large enough such that one will always detect photons regardless of the spin states. In the generalized criterion, we determine the spin to be spin-up if we detect more than $k$ photons within an integration window, otherwise we determine it is spin-down. For each $\eta$, we choose $k$ such that it maximizes the fidelity. At small $\eta$, this new criterion degenerates to our previous criterion. In Ref. [23], we showed that under this generalized criterion, the spin readout fidelity is given by

$$F = \frac{1}{2} + \frac{1}{2} \cdot \max_T \left\{ \sum_{j=0}^{M} \frac{1}{j!} \cdot \left\{ \left[ N_\downarrow (T) \right]^j e^{-N_\downarrow (T)} - \left[ N_\uparrow (T) \right]^j e^{-N_\uparrow (T)} \right\} \right\}, \tag{29}$$

where $M = \left\lfloor \dfrac{N_\uparrow (T) - N_\downarrow (T)}{\ln\left[ N_\uparrow (T) \right] - \ln\left[ N_\downarrow (T) \right]} \right\rfloor$ is the threshold photon number that gives the optimal fidelity, and $\lfloor x \rfloor$ indicates the largest integer that is not greater than $x$. We numerically calculate $N_\uparrow (T)$ and $N_\downarrow (T)$ following the same expressions as given by Eq. (27) and (28).